\documentclass[twocolumn, deluxetables]{aastex631}
\usepackage{amsmath}
\usepackage[caption=false]{subfig}
\usepackage{multirow}
\usepackage{graphicx}
\graphicspath{{./}{figures/}}

\shorttitle{XRISM Spectroscopy of GRS~1915$+$105}
\shortauthors{Miller et al.}
\begin{document}

\title{XRISM Spectroscopy of the Stellar-mass Black Hole GRS~1915$+$105}

\author[0000-0003-2869-7682]{Jon M. Miller}
\email{jonmm@umich.edu}
\affiliation{Department of Astronomy, University of Michigan, Ann Arbor, MI, 48109, USA}

\author[0000-0001-9911-7038]{Liyi Gu}
\affiliation{SRON Netherlands Institute for Space Research, Leiden, The Netherlands} %10

\author[0000-0002-7868-1622]{John Raymond}
\affiliation{Center for Astrophysics | Harvard-Smithsonian, MA 02138, USA}

\author[0000-0003-2663-1954]{Laura Brenneman}
\affiliation{Center for Astrophysics | Harvard-Smithsonian, MA 02138, USA}

\author[0000-0001-5802-6041]{Elena Gallo}
\affiliation{Department of Astronomy, University of Michigan, Ann Arbor, MI, 48109, USA}

\author[0000-0003-3105-2615]{Poshak Gandhi}
\affiliation{School of Physics \& Astronomy, University of
  Southampton, Southampton SO17 1BJ, UK}

\author{Timothy Kallman}
\affiliation{NASA / Goddard Space Flight Center, Greenbelt, MD 20771, USA}

\author{Shogo Kobayashi}
\affiliation{Department of Physics, Tokyo University of Science, 1-3 Kagurazaka, Shinjuku-ku, Tokyo 162-8601, Japan}

\author[0000-0001-7557-9713]{Junjie Mao}
\affiliation{Department of Astronomy, Tsinghua University, Haidian DS 100084, Beijing, People's Republic of China}

\author[0000-0002-5359-9497]{Daniele Rogantini}
\affiliation{Department of Astronomy and Astrophysics, University of Chicago, Chicago, IL 60637, USA}

\author[0000-0001-8195-6546]{Megumi Shidatsu}
\affiliation{Department of Physics, Ehime University, Ehime 790-8577, Japan} 

\author[0000-0001-7821-6715]{Yoshihiro Ueda}
\affiliation{Department of Astronomy, Kyoto University, Kyoto 606-8502, Japan} 

\author[0000-0002-7129-4654]{Xin Xiang}
\affiliation{Department of Astronomy, University of Michigan, Ann Arbor, MI, 48109, USA}

\author[0000-0002-0572-9613]{Abderahmen Zoghbi}
\affiliation{Department of Astronomy, The University of Maryland, College Park, MD 20742, USA}
\affiliation{HEASARC, Code 6601, NASA/GSFC, Greenbelt, MD 20771, USA}
\affiliation{CRESST II, NASA Goddard Space Flight Center, Greenbelt, MD 20771, USA}




\begin{abstract}
GRS~1915$+$105 was the stellar-mass black hole that best reproduced
key phenomena that are also observed in Type-1 active galactic nuclei.
In recent years, however, it has evolved to resemble a Type-2 or
Compton-thick AGN.  Herein, we report on the first XRISM observation
of GRS~1915$+$105.  The high-resolution Resolve calorimeter spectrum
reveals that a sub-Eddington central engine is covered by a layer of
warm, Compton-thick gas.  With the obscuration acting as a
coronagraph, numerous strong, narrow emission lines from He-like and
H-like charge states of Si, S, Ar, Ca, Cr, Mn, Fe, and Ni dominate the
spectrum.  Radiative recombination continuum (RRC) features are also
observed, signaling that much of the emitting gas is photoionized.
The line spectrum can be fit by three photoionized emission zones,
with broadening and bulk velocities suggestive of an origin in the
outer disk atmosphere and/or a slow wind at $r \simeq
10^{6}~GM/c^{2}$.  The Fe~XXV He-$\alpha$ and Fe XXVI Ly-$\alpha$
lines have a broad base that may indicate some emission from $r \sim
3\times 10^{3}~GM/c^{2}$.  These results broadly support a picture
wherein the current state in GRS~1915$+$105 is due to obscuration by
the irradiated outer disk.  This could arise through disk thickening
if the Eddington fraction is higher than inferred, but it is more
likely due to a warped, precessing disk that has brought the outer
disk into the line of sight.  We discuss the strengths and weaknesses
of this interpretation and our modeling, and possible explanations of
some potentially novel spectral features.
\end{abstract}


\section{Introduction}

GRS~1915$+$105 is a stellar-mass black hole that acts as a bright,
local laboratory for understanding accretion physics in active
galactic nuclei (AGN).  It is sometimes referred to as a
``microquasar'' for its occasional, apparently superluminal jet
velocity, and the variety of continuous and discrete radio jets that
it launches (e.g., \citealt{mirabel1994}; for a review, see
\citealt{fenderbelloni2004}).  Recently, the impact of these jets on
the local interstellar medium has been realized \citep{motta2025},
potentially revealing scaled versions of the jet cavities that massive
black holes inflate in diffuse cluster gas (see, e.g.,
\citealt{fabian2003}).

GRS~1915$+$105 is also known for a broad array of at least 12
variability patterns that may reflect particular configurations of the
disk, corona, and jet \citep{belloni2000}.  The character of disk
winds in GRS~1915$+$105 appears to vary with spectral state and the
degree of radio jet activity (see, e.g., \citealt{neilsen2009}, but
some wind activity may persist in all states \citep{neilsen2018}.
However, in soft, sub-Eddington, disk-dominated states, GRS~1915$+$105
launches disk winds with velocities as high as $v = 0.03c$
\citep{miller2016}, similar to ``ultra-fast outflows'' in Seyferts and
quasars (for a review, see \citealt{gmc2023}; also see, e.g.,
\citealt{tombesi2010}, \citealt{xiang2025}).

In recent years, the character of GRS~1915$+$105 has radically
changed.  It no longer mimics unobscured Seyferts or quasars; rather,
it more closely resembles highly obscured Seyfert-2 AGN or even
Compton-thick AGN (CTAGN; $N_{H}\geq 1.6\times 10^{24}~{\rm
  cm}^{-2}$).  Chandra grating spectra in this ``obscured state''
reveal that the internal column may be Compton-thick, and the result
of failed disk winds \citep{miller2020}.  JWST observations of
GRS~1915$+$105 in this state further confirm a large obscuring region,
and a central source that continues to accrete at 5--30\% of its
Eddington limit \citep{gandhi2024}.  Extensive monitoring with the
Neil Gehrels Swift Observatory suggests a median internal column
density of $N_{H} = 7\times 10^{23}~{\rm cm}^{-2}$
\citep{balakrishnan2021}.

GRS~1915$+$105 is a wide binary system, with a period of of
P$=33.85\pm 0.16$~days \citep{steeghs2013}.  In other wide binaries,
super-orbital periods are sometimes observed that likely result from
the precession of a warped accretion disk (e.g.,
\citealt{brumback2020}, \citealt{kosec2023}).  Recent changes in the
position angle of the radio jet in GRS~1915$+$105 may indicate that
its disk is also warped and precessing \citep{rodriguez2025}.
In the context of AGN unification models \citep{antonucci1993} --
which largely ascribe observational differences to viewing angle --
such a geometrical change would be a particularly apt means of
changing a ``microquasar'' into a micro-CTAGN.

The other key properties of GRS~1915$+$105 are known well, but are
continually refined owing to its importance.  Radio parallax
measurements now place the source at a distance of $d=9.4\pm 0.6$~kpc
\citep{reid2023}.  When coupled with prior radial velocity studies,
this distance gives a black hole mass of $M = 11\pm 2~M_{\odot}$, an
inclination of $\theta = 64\pm 4$~degrees, and jet velocities between
$0.68 \leq v/c \leq 0.91$ (\citealt{reid2023}; also see
\citealt{reid2014}).  Owing to the extreme nature its jets and the
possibility that they are powered by the spin of the black hole (e.g.,
\citealt{bz77}), considerable effort has also been devoted to measuring
this aspect of GRS~1915$+$105.  The higher radiative efficiency of
accretion onto a highly spinning black hole may be important to the
obscured state of GRS~1915$+$105, as the enhanced radiative output may
enable thickening in the outer disk or the formation a radiation-induced
warp (e.g., \citealt{pringle1996}).

Initial fits to X-ray spectra obtained with NuSTAR measured a black
hole spin parameter of $a = 0.98\pm 0.01$ (where $a = cJ/GM^{2}$, and
$J$ is the angular momentum) and an inclination of $\theta = 71\pm
1$~degrees \citep{miller2013}.  New fits to multiple spectra with
improved models measure a black hole spin parameter of $a = 0.98\pm
0.02$ and an inclination of $i = 60\pm 8$~degrees \citep{draghis2024}.
It is particularly notable that the fully independent reflection
modeling and radio parallax measurements yield commensurate
inclination values.  Using older values for the black hole mass,
distance, and inclination, fits to the X-ray continuum from the disk
measure black hole spin parameters of $a\geq 0.98$
\citep{mcclintock2006} and $a\simeq 0.7$ \citep{middleton2006}.  The
latter of these matches the value of $a=0.71\pm 0.03$ derived from
associating 67~Hz quasi-periodic oscillations (QPOs) observed in
GRS~1915$+$105 with nodal precession \citep{motta2024}.  This value is
twice higher than spins derived in other stellar-mass black holes
using the same method.

This work summarizes an initial analysis of the first Resolve
microcalorimeter X-ray spectrum of GRS~1915$+$105.  Section 2
describes the observation details and data reduction procedures.  The
analysis and results are presented in Section 3.  We discuss these
results, important caveats, and questions for future work in Section
4.  Finally, Section 5 restates our most important findings.

\section{Observations and Data Reduction}
GRS~1915$+$105 was observed with the X-ray Imaging and Spectroscopy
Mission (XRISM; \citealt{tashiro2024}) on 2024 October 17, starting at
01:40:18 UTC, with an associated observation identifier of
201068010. The total observation spanned 66.7~ks.  XRISM carries two
instruments: Resolve, an X-ray microcalorimeter \citep{ishisaki2022},
and Xtend, a CCD imager and spectrometer \citep{hayashida2018}).  At
the time GRS~1915$+$105 was observed, the Resolve gate valve was
closed, truncating the spectrum below 1.6 keV. Nevertheless, Resolve
achieves a resolution of just 4.5~eV, extends smoothly to an energy of
17.6~keV, and offers fundamentally new insights into black hole
accretion.  For this reason, the Resolve data are the sole focus of
this paper.

The data were reduced using the tools in HEASOFT version 6.34, and the
associated calibration files.  After accounting for efficiencies in
low-Earth orbit and other filtering, the net exposure time was
35.7~ks.  High-resolution primary (or, ``Hp'') events were from the
full 6x6 Resolve array, excluding the calibration pixel (36) and two
pixels that sometimes give anomalous readings (11 and 27).  Light
curves generated from the resulting event list do not show strong
variability, enabling an analysis of the full time-averaged spectrum.
A spectrum was then extracted from the event list using standard
0.5~eV spectral bins, and the tools \texttt{rslmkrmf} and
\texttt{xaarfgen} were used to produce the required redistribution
matrix file (rmf) and ancillary response file (arf).  We elected to
generate a ``large'' response matrix as the low energy continuum is
not visible in GRS~1915$+$105.

\section{Analysis and Results} 

\subsection{Model Construction}
The time-averaged Resolve spectrum of GRS~1915$+$105 was analyzed
using SPEX version 3.08.02 \citep{kaastra1996}.  The closed gate valve
truncates Resolve spectra below 1.6~keV, and the spectrum approaches
the background above 11.6~keV, so fits were made to the 1.6--11.6~keV
band.  All fits minimized a Cash statistic \citep{cash1979}.  Prior to
fitting, the spectrum was binned according to the ``optimal'' binning
algorithm of \cite{kaastra2016}.  We utilized the ``pion''
photoionization model to describe the emitted spectrum
(\citealt{miller2015}, \citealt{mehdipour2016}), which is native with
SPEX.  When utilized within SPEX, pion has the advantages of (1)
illuminating the diffuse gas with the best-fit continuum at each step
of the fitting process, and (2) layering different zones so that outer
regions see a properly attenuated ionizing continuum.  To capture the
full ionizing band for pion calculations, luminosities were
extrapolated to the 0.0136--13.6~keV band.

Figure 1 shows the spectrum over the 2--10~keV pass band, and Figure 2
expands the spectrum in 2~keV slices.  The spectrum is dominated by
extremely strong and narrow emission lines.  In the case of the Fe~XXV
He-$\alpha$ complex between 6.6--6.7~keV, for instance, the lines are
8--10 times stronger than the local continuum (see Figure 3).  This
immediately points to a complex physical environment, wherein
obscuration and other effects must be important.  A completely unique
characterization of the spectrum may not be possible.  However, we
constructed a plausible model based on key physical considerations and
important precedents:\\

\noindent$\bullet$ The prior Chandra observations in the obscured
state implied a luminosity as high as $L \simeq 10^{38}~{\rm erg}~{\rm
  s}^{-1}$ (0.3-10.0~keV), or $\lambda_{Edd} \simeq 0.07$
\citep{miller2020}.  In this regime, the continuum should be a mixture
of hot accretion disk and power-law components, so we allowed for both
via the ``dbb'' and ``pow'' components.  To avoid issues with the
power-law in photoionization modeling, the ``pow'' component was bent
to zero flux at low and high energy using two ``etau'' components.
The associated free continuum parameters include the power-law index
($\Gamma$) and flux normalization, the disk peak temperature (about
twice the temperature at the innermost orbit), and the disk flux
normalization.\\

\noindent$\bullet$ The strength of the observed spectral lines helps
to constrain the luminosity of the continuum components.  Their
luminosity must be much higher than is naively inferred, in order to
produce the lines.  Plausible columns of neutral gas predict a strong,
sharp, Fe~K edge that is not observed.  This signals that the column
must be warm, rather than cold and completely neutral.  We modeled
obscuration of the continuum using the ``hot'' model, leaving the gas
column density, temperature, and covering fraction free to vary.  We
fixed an internal rms velocity of $\sigma = 1000~{\rm km}~{\rm
  s}^{-1}$ as this value matches the data but is otherwise poorly
constrained.\\

\noindent$\bullet$ The presence of He-like and H-like RRCs in the
spectrum (see Figures 1, 2, and 3) requires a prominent role for
photoionization. The need for photoionization is also evident in the
ratio of the lines in the Fe~XXV complex between 6.6--6.7~keV, and
other He-like complexes.  We therefore allowed multiple ``pion''
emission zones to vary within our model, finding that {\it at least}
three are required to fit the data (see Figure 4).  Note that the
continuum from the central engine ``feeds into'' each pion component.
In our final model, the zones are layered assuming that ionization
falls with radius.  The free parameters for each pion emission zone
include the gas column density (${\rm N}_{\rm H}$), the log of the gas
ionization parameter ($\xi$), the rms velocity of the gas within the
zone ($\sigma$), the bulk velocity shift of the zone ($v$), and the
covering factor of the emitting gas ($\Omega$, measured as
$\Omega/4\pi$).
 
In preliminary fits, the pion zones all adopted very high columns,
covering factors significantly above unity, and super-solar elemental
abundances.  High colums are possible, but covering factors above
unity are unphysical, and prior studies do not support extreme
abundances \citep{keshet2025}.  The unphysical covering factors likely
resulted from the gas density in each zone being left at minimum
default values.  Since recombination and collisional excitation (which
must contribute at some level) depend on $n^{2}$, higher density
values will produce stronger line emission.  Simple estimates and
experimentation with fits suggest that a value of $n = 3\times
10^{10}~{\rm cm}^{-3}$ is sufficient to produce the observed lines,
and compatible with expectations for the outer accretion disk (e.g.,
\citealt{ss73}).  This density was fixed in all three pion zones, and
a limit of $\Omega/4\pi \leq 1$ was also enforced in each zone.\\

\noindent$\bullet$ The ratios of specific lines differ slightly from
simple expectations, regardless of the emission mechanism.  For
instance, the two components of the H-like Fe~XXVI line (nominally at
6.95 and 6.97~keV) should have a 1:1.7 flux ratio; instead, they are
nearly equal in strength.  This suggests that resonant scattering in a
hot medium removes some photons to alter the flux ratio, similar to
effects observed in the Hitomi spectrum of the Perseus Cluster
\citep{hitomi2016}.  We model this effect by covering all components
with a ``hot'' component in SPEX, allowing the gas column density
(${\rm N}_{\rm H, RS}$), covering factor ($f_{cov, RS}$), and
temperature to vary.\\

\noindent$\bullet$ To model the neural Fe~K emission in the spectrum
via illumination of distant cold gas, we imported the ``mytorus''
function \citep{my2009} and allowed for dynamical broadening via the
``spei'' component in SPEX \citep{speith1995}.  ``Mytorus'' includes
the key atomic structure necessary for calorimeter resolution,
including K$_{\alpha,1}$ and K$_{\alpha,2}$ lines, as well as
$K_{\beta}$ emission.  The key parameters of ``mytorus'' include the
gas column density, the power-law index of the incident radiation
(fixed to $\Gamma = 2.2$ in our fits for simplicity), the inclination
of the emitting gas, a velocity shift (set to zero in our fits), and a
line flux normalization.  For simplicity and because we anticipate the
neutral line arises in the optically thick disk, we fixed the column
density within ``mytorus'' to a value of ${\rm N}_{\rm H} = 1.6\times
10^{24}~{\rm cm}^{-2}$. \\

``Spei'' is very sophisticated and allows for many specific
realizations; we assumed an emissivity of $q=3$ (where $J\propto
r^{-q}$ and $q=3$ corresponds to a flat disk), and measured the inner
emission radius and inclination of the gas (the ``mytorus''
inclination parameter was coupled to this one).  The outer emission
radius was arbitrarily fixed to $r_{out} = 10^{6}~GM/c^{2}$.  In
total, then, the neutral Fe~K complex was modeled using three
parameters: the line flux normalization, inner radius, and
inclination.  This scheme follows fits to the neutral Fe~K$_{\alpha}$
and K$_{\beta}$ lines in Resolve spectra of the Seyfert-1.5 AGN
NGC~4151 \citep{miller2024}.\\

\noindent$\bullet$ It is apparent that there may be a broad base to
the Fe~XXV~He-$\alpha$ complex, and Fe~XXVI~Ly-$\alpha$
doublet. Simple Gaussian functions were added to model any broad line
flux. Within SPEX, Gaussian parameters include the central energy,
FWHM, and flux normalization of the line.  The Fe~XXV feature was
constrained to have central energy between 6.65--6.70~keV, and a width
less than FWHM$\leq0.3$~keV ($5700~{\rm km}~{\rm s}^{-1}$).  The
Fe~XXVI feature was constrained to have a central energy between
6.95--6.70~keV, and a width less than FWHM$\leq 0.1$~keV ($1800~{\rm
  km}~{\rm s}^{-1}$).\\

\noindent$\bullet$ All of these components are seen through the
interstellar medium (ISM), which we modeled with a separate fixed
``hot'' component (with the gas temperature sit to a minimum value of
$kT = 1\times 10^{-6}$~keV), assuming a covering factor of unity.

The full model can be written as:\\

\noindent ${\rm N}_{\rm H, ISM}\times [{\rm N}_{\rm H,CE} \times {\rm N}_{\rm H, RS} \times ({\rm dbb} + {\rm pow}) + {\rm N}_{\rm H, RS} \times ({\rm pion}_1 + {\rm pion}_2 + {\rm pion}_3 + {\rm spei}\times {\rm mytorus})]$,\\

\noindent where ${\rm N}_{\rm H, ISM}$, ${\rm N}_{\rm H,CE}$, and
${\rm N}_{\rm H, RS}$ refer to components accounting for absorption
and scattering in the ISM, absorption and scattering of continuum
emission from the central engine, and resonant scattering of line
emission, respectively.  The results of fitting this model to the
time-averaged Resolve spectrum of GRS~1915$+$105 are detailed in Table
1.

\subsection{Oscillations in the Fe XXVI RRC?}

RRCs arise through the recombination of hot electrons onto atoms.  The
energy of the emitted photon is the energy of the captured electron,
plus the (negative) ionization threshold of the atom.  A smooth
distribution of electron energies therefore gives rise to a smooth
flux decline above the threshold energy.  However, a series of eight
or more features -- spaced by a regular interval of 30~eV -- are
evident in the spectrum of GRS~1915$+$105 (see Figure 5).  If the
oscillations are real, they are potentially consistent with
recombination from electrons in Landau levels in a highly magnetized
gas (e.g., in a corona above the disk).  However, this explanation is
not unique.

To explore these feature, a new version of ``pion'' was developed for
this analysis and included in the latest publicly released version of
SPEX (version 3.08.02).  Within pion, the magnetic field of the gas
can be determined through such oscillations and/or through Zeeman
splitting effects.  We made a number of additional fits in the
8.4-10.0~keV band, activating the magnetic features only in the most
highly ionized ``pion'' component (pion$_1$ in Table 1). In total, the
features require three additional free parameters: the electron
magnetic field, the phase of the oscillations, and the amplitude of
the oscillations.

\subsection{Results}

The results of fits to the Resolve spectrum of GRS~1915$+$105 with the
(non-magnetic) model described above are detailed in Table 1, and
shown in Figures 1--5.  The model achieves a Cash statistic of $C =
3891$ for $\nu = 2933$ degrees of freedom.  While this does not
represent a formally acceptable fit, it is clear in Figures 1--5 that
all of the strong emission lines are fit well, that key line ratios
are largely reproduced, the evident red-shifts are accurately modeled,
the predicted RRCs match the data, and the continuum is modeled well.

In the obscured state of GRS~1915$+$105, the continuum has been
particularly difficult to determine.  Prior fits to Chandra spectra of
the obscured state did not require a disk component
\citep{miller2020}, likely owing to the limited sensitivity of the
gratings spectra.  The continuum measured with Resolve is strongly
dominated by the disk component, characterized by a temperature of $kT
= 3.53^{+0.06}_{-0.05}$~keV.  The best fit with this model does not
strongly require a power-law, so we fixed a value of $\Gamma = 2.5$
and measured a weak flux with large upper limits.  The measured
continuum gives a total continuum luminosity of $L = 8.8\pm 0.6\times
10^{37}~{\rm erg}~{\rm s}^{-1}$ in the 0.0136--13.6~keV band, or
$\lambda_{Edd.} = 0.062$.

Within SPEX, the disk component temperature is measured at the radius
of peak emissivity, which exceeds the temperature at the innermost
stable circular orbit (ISCO) by a factor of approximately $\simeq2$.
A value of $kT \simeq$1.7-1.8~keV is likely more appropriate for
comparisons to XSPEC models \citep{arnaud1996}, which measure
temperatures at the ISCO.  Even this temperature is likely higher than
the emitted temperature, owing to scattering in gas along the line of
sight.  In a disk-dominated, unobscured state of GRS~1915$+$105,
Chandra measured an inner disk temperature of $kT = 1.6$~keV, and a
power-law index of $\Gamma = 2.8$ \citep{miller2016}.  The broad
similarity of these values to those measured with Resolve -- despite
very different obscuration profiles -- suggests that Resolve observed
GRS~1915$+$105 in an ordinary, sub-Eddington, ``thermal-dominant'' or
``high soft'' state, rather than a super- or hyper-Eddington state
dominated by a geometrically thick inner disk.  (For a review of black
hole states, see, e.g., \citealt{rm2006}.)

As noted above, the obscuring column cannot be neutral, but gas
temperatures of even $kT = 0.1$~keV are easily excluded.  The model
indicates that a warm, $kT = 0.002^{+0.008}_{-0.001}$~keV,
Compton-thick column of $N_{H} = 3.0^{+0.3}_{-0.3}\times 10^{24}~{\rm
  cm}^{-2}$ covers most of the central engine ($f_{cov} = 0.92\pm
0.02$).  The warm gas supplies the column required to match the
observed continuum, while obscuring a central engine that is luminous
enough to produce the observed emission lines.  It also matches subtle
features in the spectrum just above the neutral Fe~K edge, where
structure is expected owing to edges from low charge states of Fe.

Using a simple $T\propto r^{-3/4}$ scaling and the inferred inner disk
temperature, the accretion disk would have a temperature of $kT =
0.002$~keV at $r \simeq 1.2\times 10^{5}~GM/c^{2}$.  The obscuring gas
likely has something to do with the accretion disk itself, since the
column is optically thick; however, at least some of the gas might
instead be attributed to a warm disk atmosphere above a disk at even
larger radii.  It is notable that Keplerian orbital velocity at this
radius is $\sigma = 840~{\rm km}~{\rm s}^{-1}$, close to the rms value
of $\sigma = 1000~{\rm km}~{\rm s}^{-1}$ that matches the data well.
In short, the nature of the obscuration is consistent with the
outer disk occulting the central engine in GRS~1915$+$105.

These radii are not compatible with the width of the neutral
Fe~K$_{\alpha}$ line.  The observed blending of the two line
components and the absence of an observable line shift require a
combination of a smaller radius and a low inclination.  In broad
terms, the neutral Fe~K$_{\alpha}$ line originates 1--2 orders of
magnitude closer to the central engine than the obscuring gas does
(see Table 1).  The low inclination of the neutral Fe~K$_{\alpha}$
line, $i = 5^{+4}_{-2}$~degrees, is very different from that of the
binary system ($i = 64\pm 4$~degrees, \citealt{reid2023}), the
innermost disk ($i= 60\pm 8$~degrees, \citealt{draghis2024}), and the
jet ($i = 66\pm 2$~degrees, \citealt{fender1999}).  Indeed, if the
disk is flat and the central engine can be treated as a point source,
each annulus should contribute progressively less emission, and we
should expect to see a broad, indistinct line.  Sharp lines from
specific radii -- e.g., from the broad line region or inner face of
the torus in AGN \citep{miller2024} -- require a geometry that adds
more emitting area per radius than a truly flat disk can produce.  These
conditions needed to produce the neutral Fe~K$_{\alpha}$ line are
broadly consistent with a warp that first departs from a flat disk at
$r \simeq 10^{3-4}~GM/c^{2}$, bringing the outer disk into the line of
sight by $r \simeq 10^{5-6}~GM/c^{2}$.

The parameters of all three pion zones are very well determined.
Figure 4 shows how the components contribute to the flux observed in
the Fe~XXV He-$\alpha$ and Fe~XXVI Ly-$\alpha$ lines. The line widths
and shifts are likely the best guide to the location of the emitting
gas. Although the gas in each zone is highly ionized, the zones
exhibits remarkably low internal rms velocities and low bulk
velocities.  None of the measured rms velocities and bulk shifts
exceed $v \geq 350~{\rm km}~{\rm s}^{-1}$.  Given the high inclination
of the binary system, our line of sight likely captures most of the
orbital velocity of the gas.  In principle, the gas may have a large
vertical velocity (perpendicular to our line of sight), and continual
acceleration could lead to even higher velocities over many scale
heights.  If the measured gas velocities are indicative of Keplerian
orbital speeds, the gas must lie at $r \geq 7\times 10^{5}~GM/c^{2}$.
This indicates that the lines are produced just outside of the region
that obscures the central engine, likely in a disk atmosphere (and/or
the base of a wind).

The Gaussians added to the model to describe the broad base of the
Fe~XXV He-$\alpha$ an Fe~XXVI Ly-$\alpha$ lines are well constrained.
The width of broad Fe~XXV line hits the imposed limit of FWHM$\leq
0.3$~keV, or $\sigma \leq 5700~{\rm km}~{\rm s}^{-1}$.  The width of
the broad Fe~XXVI Ly~$\alpha$ line is measured to be $\sigma =
700_{-200}^{+100}~{\rm km}~{\rm s}^{-1}$.  The Fe~XXV line could arise
as close as $r \simeq 3\times 10^{3}~GM/c^{2}$ -- broadly compatible
with the neutral Fe~K$_{\alpha}$ emission line -- and potentially
indicating that the leading edge a putative warp could be a complex
geometry that churns up a combination of neutral and ionized gas.

Using the most recent binary system parameters, the semi-major axis of
the components in GRS~1915$+$105 is $a \simeq 7\times 10^{12}~{\rm
  cm}$, or $a = 4.2\times10^{6}~GM/c^{2}$.  Using the approximations
in \cite{eggleton1983}, the radius of the Roche lobe around the black
hole is likely $r_{roche} = 4.4\times 10^{12}~{\rm cm}$, or $r_{roche}
= 2.7\times 10^{6}~GM/c^{2}$.  In most circumstances, the accretion
disk is expected to fill about two-thirds of its Roche lobe, which
suggests that the outer disk extends to $r_{out} = 1.8\times
10^{6}~GM/c^{2}$.  These estimates confirm that the disk is just large
enough to produce the strong, narrow, ionized emission lines that we
have observed.

Although the broadening and shifts observed in each photoionized
emission zone are small, they are significantly different.  It is not
clear if this reflects radial or vertical stratification within a disk
atmosphere or wind base, or if the differences indicate axial
asymmetry.  For instance, pion$_1$ is significantly blue-shifted ($v_1
= -90^{-10}_{+10}~{\rm km}~{\rm s}^{-1}$), while $pion_2$ is
significantly red-shifted ($v_2 = 342^{+6}_{-7}~{\rm km}~{\rm
  s}^{-1}$), and $pion_3$ is hardly shifted at all ($v_3 =
28^{+9}_{-6}~{\rm km}~{\rm s}^{-1}$).  It is at least possible that
these shifts reflect axial asymmetry brought about by a large-scale
warp in the outer disk.

Thermal winds can be launched outside of the Compton radius, given by
${\rm R}_{\rm C} = 1\times 10^{10}\times ({\rm M}_{BH}/{\rm
  M}_{\odot})\times {\rm T}_{\rm C,8}^{-1}$ (where ${\rm T}_{\rm C,8}$
is the Compton temperature in units of $10^{8}$~K; \citealt{bms83}).
Approximating the Compton temperature as the disk temperature, ${\rm
  R}_{\rm C} = 1.6\times 10^{5}~GM/c^{2}$, inside the radius at which
the photoionized lines are produced.  It is therefore plausible
that the narrow lines trace a thermal wind.  Indeed, the large
inferred distance from the central engine, high ionizations, and low
velocities are all broadly consistent with numerical simulations of
thermal winds (see, e.g., \citealt{higginbottom2017}), but the
inconsistent shifts may again point to potential axial asymmetry.  The
broad line that underlies the narrow Fe~XXV He-$\alpha$ complex may
not be part of a thermal wind, given an origin at $r\simeq 3\times
10^{3}~GM/c^{2}$.

The resonant scattering medium that is inferred via the Fe~XXV
He-$\alpha$ and Fe~XXVI Ly-$\alpha$ line ratios likely represents a
hot, fast, ionized wind.  The inferred temperature, $kT =
48^{+5}_{-6}$~keV, is far above the escape speed in the outer disk.
Moreover, the best-fit velocity of the inferred medium is $v =
-610^{-30}_{+30}~{\rm km}~{\rm s}^{-1}$, largely set by the ratio of
the Fe~XXVI Ly-$\alpha$ line.  Given that the line of sight to the
central engine in GRS~1915$+$105 is now blocked by the outer disk (or,
skims along the surface of the disk), most of the velocity in this
component may not be detected.

Our model requires only a few modest departures from solar abundances.
The data weakly prefer 20--40\% abundance enhancements for Si, Ar, Fe,
and Ni, relative to solar values.  However, the data require larger
and more significant enhancements for other elements: ${\rm A}_{\rm
  Ca} = 1.54^{+0.03}_{-0.14}$, ${\rm A}_{\rm Cr} = 2.9^{+0.2}_{-0.4}$,
and ${\rm A}_{\rm Mn} = 2.8^{+0.2}_{-0.4}$.  Mn is the only element
contributing strong lines with an odd atomic number (25), so its
abundance measurement may be the most important. The value indicated
by the Resolve data is lower than the value of ${\rm A}_{\rm Mn} =
4\pm 1$ reported by \cite{keshet2025}, but consistent within errors.
The lower abundance that we have measured brings the overall abundance
pattern in GRS~1915$+$105 closer to predicted SN yields (for a review,
see \citealt{nomoto2013}), and strengthens evidence that the
progenitor event deposited some elements into the companion star.

\subsection{The Fe XXVI RRC}

Figure 5 shows the results of fits to the 8.4--10.0~keV band using the
model detailed in Table 1, and using an alternative model with
magnetic field effects activated.  Over this pass band, the reference
model delivers a Cash statistic of $C = 498$ for $\nu = 393$ degrees
of freedom.  With magnetic effects activated in the most highly
ionized ``pion'' emission zone (pion$_1$ in Table 1), the Cash
statistic improves to $c = 446$ for $\nu = 390$.  Were these values
strictly equivalent to $\chi^{2}$ statistics, this improvement would
be significant at the $5.5\sigma$ level of confidence.

The best-fit magnetic field is measured to be ${\rm B} = 2.57\pm
0.06\times 10^{9}$~Gauss.  The phase of the observed oscillations is
measured to be $\phi = 0.13^{+0.08}_{-0.03}$, and the amplitude is
measured to be $A = 1.9\pm 0.4$.  This field is extremely high,
commensurate with the surface magnetic field inferred in accreting
millisecond X-ray pulsars \citep{patruno2021}.  The putative field
would have to exist in a hot, electron-dominated corona that may
differ from the underlying accretion disk.  However, assuming that
these features are also produced in the outermost accretion disk, the
implied magnetic field is several orders of magnitude greater than
equipartition between magnetic pressure and gas pressure
(\citealt{ss73}; also see \citealt{miller2016}). A field this high
would easily disrupt the disk, and it is not clear how an
equipartition field that might arise in the disk could be amplified to
such a degree in a blanketing atmosphere or corona; in summary, the
features may not be due to magnetic effects (also see Section 4.5, below).

Moreover, the apparent oscillations are quite sharp.  Whatever the
exact nature of the gas, separations of 30~eV would
be completely blurred out at radii less than $r \leq 1.1\times
10^{5}~GM/c^{2}$ and so much larger radii -- $r\simeq 1\times
10^{6}~GM/c^{2}$ -- are likely.  At this radius, however, the corona
would be unbound, and could represent an ionized outflow.  The
distinct velocities are qualitatively similar to the wind observed in
PDS~456, and so are the velocities \citep{hagino2025}.  The lines
cannot be dominated by H-like Fe~XXVI, because the 18~eV spacing of
the line components is incompatible with the data. Rather,
intermediate charge states would be required, and velocities up to
$v=0.34c$.  If the wind is launched verticaly, the true velocities
would be even higher.

%---------------------------------------------------------

\section{Discussion}
We have analyzed the first XRISM/Resolve spectrum of the stellar-mass
black hole GRS~1915$+$105.  The source was captured in its
``obscured'' state.  Our models suggest that the central engine
remains in a disk--dominated, sub-Eddington accretion mode, but is
largely occulted by the outer accretion disk.  This situation is
qualitatively similar to highly obscured and Compton-thick AGN,
wherein the central engine is obscured by the torus at the outer edge
of the accretion flow.   However, there are important differences,
including that the torus is cold enough to retain dust and molecules,
whereas the outer disk in X-ray binaries is not.  In this section, we
discuss the nature of the ``obscured'' state in more detail,
considering multi-wavelength observations and analogous sources.  We
also discuss the shortcomings of our model, alternative emission
mechanisms that may contribute to the strong emission line spectrum,
and comment on apparent structure in the Fe~XXVI~RRC.

\subsection{The nature of the obscured state}

Our model describes the Resolve spectrum of GRS~1915$+$105 in terms of
a standard, sub-Eddington inner accretion flow that is obscured by
warm, Compton-thick gas.  The temperature of the obscuring medium is
consistent with the outer accretion disk.  The level of obscuration
that is measured with Resolve exceeds values measured using Swift and
Chandra (e.g., \citealt{balakrishnan2021}, \citealt{miller2020}).
While this could indicate that the central engine in GRS~1915$+$105 is
now even more deeply buried than it was in the recent past, the
differences are more likely attributable to different levels of
sensitivity and the modeling choices that follow.  The role of warm
gas absorption and the shortcomings of neutral obscuration were likely
not apparent in spectra with more limited sensitivity and resolution.

A super-Eddington accretion rate could cause the inner accretion disk
to assume a ``funnel'' geometry \citep{king2001}, blocking some of the
ionizing flux and potentially accounting for some aspects of the
``obscured'' state.  This configuration remained a possibility based
on Chandra gratings spectra obtained in the ``obscured'' state, which
did not require an accretion disk.  The much higher sensitivity
achieved even in this short Resolve spectrum does require a disk, but
the disk temperature is typical of bright,  sub-Eddington soft
states (e.g., \citealt{hb2005}, also see \citealt{rm2006}).

The best explanation of the Resolve spectrum is that the ``obscured''
state is that the central engine is occulted by the outer accetion
disk.  This is indicated by the nature of the obscuring gas itself,
the radii and inclinations measured from the neutral Fe~K$_{\alpha}$
line and broad ionized lines, and the radii inferred from the narrow
ionized emission lines.  The question, then, is why the outer disk
obscures the central engine now, whereas it did not do so
historically.

Seminal theoretical treatments find that radiation pressure on the
outer accretion disk can drive a warp, leading to super-orbital
periods (e.g., \citealt{pringle1996}).  In sources that are viewed at
a high inclination angle -- just above the plane of the disk --
precession may occasionally cause the outer disk to partly or
completely occult the central engine.  The condition is that the disk
must be fairly large.  This condition is surely met in GRS~1915$+$105,
which has the longest orbital period and widest orbital separation
among black hole X-ray binaries.  The super-orbital periods and low
flux states observed in neutron star X-ray binaries such as Her X-1,
SMC X-1, and LMC X-4 are attributed to warping, precession, and
occultation (see, e.g, \citealt{brumback2020}, \citealt{kosec2023}).
The orbital periods of those systems are all less than four days, so
similar phenomena in GRS~1915$+$105 should play out over much longer
time scales.

Radio observations of GRS~1915$+$105 in 2023 reveal a 24~degree change
in the position angle of the jets since 1999, as well as a 17~degree
change in the inclination of the jets with respect to the line of
sight \citep{rodriguez2025}.  This degree of change in the broad
accretion flow is likely sufficient to bring the outer disk into the
line of sight to the central engine in GRS~1915$+$105.  Recent
observations with JWST measure an IR flux that is 10 times higher than
in the past, and very strong IR recombination lines, indicating an
enhanced role for reprocessing in the ``obscured'' state.
Importantly, the brightest Pf(6-5) line lags the continuum by the
expected light travel time from the central engine to the outer disk
\citep{gandhi2024}.  

Taken as a whole, the Resolve data, radio data, and JWST data deliver
a broadly self consistent picture: the ``obscured'' state of
GRS~1915$+$105 is due to a warped, precessing accretion disk that is
temporarily occulting the central engine.  The length of the active
period prior to the current ``obscured'' state, and the length of the
``obscured'' state itself, are likely a consequence of the fact that
GRS~1915$+$105 is a very large X-ray binary, with a long orbital
period.  While the accretion flow occults the central engine, we are
afforded an unprecedented multi-wavelength view of the outer accretion
disk atmosphere and connected winds.  Figure 6 shows a schematic
representation of a geometry that could produce the Resolve
spectrum.It is interesting to speculate that GRS~1915$+$105 may be a
persistent source, and that the central engine was merely obscured
prior to its detection in 1992 \citep{ct92}, not truly quiescent.

We note that while the multi-wavelength data and some details of the
Resolve spectrum favor a warp, a less extreme version of the ULX
scenario cannot be ruled out.  It is possible that the inner flow is
close to the Eddington limit, but buried by an even higher column than
inferred by our model.  In this case, the inner disk may not form into
a funnel, but the outer disk would be more strongly illuminated and
increase in scale height.  In principle, at least, this could cause
the outer disk in GRS~1915$+$105 to now intercept the line of sight,
rather than a warp.

\subsection{Model shortcomings}

The model detailed in Table 1 captures many key aspects of the Resolve
spectrum of GRS~1915$+$105, but it has some shortcomings.  For
instance, we have assumed that each zone has a gas density of $n =
3\times 10^{10}~{\rm cm}^{-3}$.  This density value is driven by the
need to produce strong lines while ensuring that $\Omega/4\pi \leq 1$.
However, there is no reason why each zone must have the same density
despite a range of ionizations and columns.  Moreover, a covering
factor of unity is unlikely and a modestly higher density of $n =
1\times 10^{11}~{\rm cm}^{-3}$ should drop the covering factor to
$\Omega/4\pi \simeq 0.1$.  Future modeling should explore higher fixed
density values, and/or explicitly fit for the density of each zone.

Equally importantly, the model only includes three photoionized
emission zones. The fact that the fit is not formally acceptable
indicates that a more granular approach may be needed.  For instance,
while the model is an excellent match for the strongest Fe and Ni
lines, the predicted Cr~XXIV lines are broader than the data (see
Figure 2).  A model with a larger number of velocity and ionization
pairings may enable improved fits.

Finally, the model neglects photoionized absorption.  Potential
absorption lines in this spectrum are easily neglected in favor of the
emission lines, but a more complete treatment could yield additional
physical insights.  Apart from features that dip below the continuum
in line-free regions, a number of the lines in the 8.0--10.0~keV band
(for instance) appear to have a double-peaked structure that may
indicate a role for absorption.

\subsection{Explaining the Fe~XXVI RRC}

Using a new capability within the ``pion'' model, we find that
apparent oscillatory structure in the Fe~XXVI~RRC feature can
described in terms of recombination from Landau levels in a corona
with a magnetic field of ${\rm B} = 2.57\pm 0.06\times 10^{9}$~Gauss
(see Figure 5).  This change is likely significant over the model
detailed in Table 1 at more than the $5\sigma$ level of significance.
However, the implied field strength is so extreme -- and so out of
step with theoretical predictions -- that it must be regarded with
skepticism.  Instead describing the features in terms of a
relativistic outflow poses other severe challenges.  We note that even
more exotic explanations for the features include ionized charge
states of gallium (requiring an over-abundance of ${\rm A}_{\rm Ga}=
5\times 10^{4}$), or a resonance due to polarized electrons with
aligned spins.

We investigated different means by which the oscillatory structure in
the Fe~XXVI RRC could arise through instrumental effects.  The
Resolve instrument team does not know of any instrumental features in
this range (Caroline Kilbourne, priv. comm.)  There was no solar
flaring reported during the observation; the astrophysical background
was likely not enhanced nor anomalous.  The presence of the features
does not appear to depend on the inclusion nor exclusion of pixels 11
and 27, which can give anomalous readings.

Nevertheless, we currently favor an as-yet unknown instrumental
artifact or unknown astrophysical effect.  Resolve is the first
(known) X-ray microcalorimeter to operate in the space environment for
an extended period.  There may not be any archival spectra, nor any
XRISM spectra, with similarly strong Fe XXV and Fe XXVI RRC features.
Observations of other systems may eventually reveal that the apparent
oscillations are real but not tied to magnetic effects.

\subsection{Alternative emission mechanisms}

We have modeled the data successfully in terms of photoionized
emission, but it is likely that collisional excitation and
non-equilibrium emission contribute a degree of line flux.  On their
own, these mechanisms fail to match the observed line ratios, and
particularly fail to produce RRC features that match the data (as
expected).  A cooling, condensing disk atmosphere, for instance, is
expected to produce particularly strong emission lines (see, e.g.,
\citealt{jg2001}).  To explore this possibility, we attempted a number
of fits using the ``nei'' non-equilibrium plasma model in SPEX.  For
low gas densities, very strong resonance and forbidden lines are
predicted, but intercombination lines fall short of the data.  When
the gas density reaches or exceeds $n = 10^{18}~{\rm cm}^{-3}$, the
predicted intercombination lines roughly match the data, but even then
the model falls short of the RRC features.  In our fitting
experiments, models with single ``nei'' components yielded Cash
statistics that were higher by $\Delta C \simeq 1000$, and
combinations of two or more ``nei'' components (with or without
photoionized zones) were higher by $\Delta C \simeq 500$.
Nevertheless, future efforts to model this spectrum should explicitly
consider small contributions from collisional plasmas and
non-equilibrium plasmas.\\

\subsection{Open questions}

The three photoionization zones detailed in Table 1 exhibit low but
inconsistent velocity widths and bulk shifts.  This combination is
necessary to match the profile of the strongest emission lines,
particularly the Fe~XXV He-$\alpha$ complex between 6.6--6.7~keV (see
Figure 4).  It is possible that the velocity structure reflects
emission from different faces of a warp, with geometries on the near
and far sides of the central engine contributing.  The single, short
observation that we have examined cannot test whether some of these
velocities reflect axial anisotropy.  Future XRISM observations that
obtain Resolve spectra throughout the binary period can better
determine the nature of the gas velocities.

If some of the gas velocities are driven by the binary system rather
than Keplerian orbital motions, the emission lines could originate in
other locations.  It is possible that some of the line flux could
arise on the irradiated surface of the companion star.  In addition,
some of the line flux could originate in a circumbinary disk; this
possibility is also a plausible explanation of the JWST data
\citep{gandhi2024}.  Finally, it is even possible that the very low
velocities do not reflect Keplerian motion, but only bulk motion above
the plane of the disk.  In principle, the photoionized gas could have
a cylindrical distribution above the disk, or even a polar
distribution that is more centrally concentrated.  Depending on its
extent, this geometry could be the X-ray binary equivalent of the
``narrow line region'' in AGN (see, e.g., \citealt{wang2011}), which
is observed in extended polar regions that can be orders of magnitude
larger than the torus.  Here again, future observations that can
detect changes in the central engine and line spectrum are needed to
test these possibilities.

\section{Summary and Conclusions}

The results of our analysis of the first Resolve spectrum of
GRS~1915$+$105 in its ``obscured state'' can be summarized as follows:\\

\noindent$\bullet$ The central engine in GRS~1915$+$105
is likely obscured by warm, partially covering, Compton-thick gas.\\  

\noindent$\bullet$ After accounting for this obscuration, the data are
consistent with a standard, sub-Eddington, inner accretion flow that
is dominated by a hot accretion disk.\\

\noindent$\bullet$ The high-resolution calorimeter spectrum is
dominated by strong, narrow emission lines from He-like and H-like
charge states, consistent with photoionization in the outer accretion
disk.\\

\noindent$\bullet$ It is therefore likely that the outer disk obscures
the central engine.  This could be due to axially symmetric thickening
of the outer disk in response to a higher central luminosity than we
have inferred, or due to warping and precession, similar to
super-orbital periods and low-flux phases in other wide binary
systems.  The latter explanation is potentially consistent with
observed changes in the position angle and inclination angle of the
radio jet in GRS~1915$+$105 \citep{rodriguez2025}.\\

\noindent$\bullet$ The observed emission lines likely probe the
atmosphere of the outer accretion disk, or the base of a slow wind
(but note that these may not be entirely distinct).  The measured
abundances suggest enhancements in Cr and Mn, and are broadly
consistent with models wherein such elements are deposited into the
companion star by the black hole creation event (\citealt{nomoto2013},
\citealt{keshet2025}).\\

% Acknowledgements
The authors thank the anonymous referee for helpful comments that
improved this manuscript.  JMM acknowledges helpful conversations with
Ehud Behar, James Miller-Jones, Jelle Kaastra, Caroline Kilbourne,
Frits Paerels, Daniel Proga, Luis Rodriguez, and
Sergei Trushkin.  JMM thanks Jelle de Plaa for sharing advanced
versions of SPEX, and thanks Savannah Ware, Elissa Slotkin, Sydney
Hess, and Mark Kelly for their efforts to keep NASA astrophysics
vibrant, including support for XRISM.

\bibliography{main}{}
\bibliographystyle{aasjournal}

\onecolumngrid

\begin{table}[t!]
\caption{Spectral Model Parameters}
\begin{scriptsize}
\begin{center}
\begin{tabular}{lll}
Parameter & Value & Comments\\
\tableline
disk kT (keV) & $3.53^{+0.06}_{-0.05}$ & about 2X higher than at innermost orbit\\
disk norm. ($10^{9}~{\rm cm}^{2}$) & $1.27^{+0.02}_{-0.02}$ &  -- \\
pow. $\Gamma$ & $2.5^{*}$ & fixed \\
pow. norm. ($10^{41}~{\rm ph}~{\rm s}^{-1}~{\rm kev}^{-1}$ & $7^{+93}_{-7}$ & effectively an upper limit \\
\tableline
%pion_1 is com 8
pion$_{1}$ ${\rm N}_{\rm H} (10^{22}~{\rm cm}^{-2})$ & $5.6^{+0.6}_{-0.5}$ & -- \\
pion$_{1}$ ${\rm log} \xi$ & $4.29^{+0.05}_{-0.03}$ & -- \\
pion$_{1}$ $\sigma ({\rm km}~{\rm s}^{-1})$ & $250^{+10}_{-10}$ & internal turbulent velocity \\
pion$_{1}$ v (${\rm km}~{\rm s}^{-1}$) & $-90^{-10}_{+10}$ & bulk velocity shift \\
pion$_{1}$ $\Omega$ & $1.00_{-0.01}$ \\
\tableline 
%pion_2 is com 7
pion$_{2}$ ${\rm N}_{\rm H} (10^{22}~{\rm cm}^{-2})$ & $1.7^{+0.1}_{-0.1}$ & -- \\
pion$_{2}$ ${\rm log} \xi$ & $3.78^{+0.03}_{-0.03}$ & -- \\
pion$_{2}$ $\sigma ({\rm km}~{\rm s}^{-1})$ & $90^{+10}_{-10}$ & internal turbulent velocity \\
pion$_{2}$ v (${\rm km}~{\rm s}^{-1}$) & $342^{+6}_{-7}$ & bulk velocity shift \\
pion$_{2}$ $\Omega$ & $1.0_{-0.03}$ \\
\tableline
%pion3 is com 12
pion$_{3}$ ${\rm N}_{\rm H} (10^{23}~{\rm cm}^{-2})$ & $3.3^{+0.3}_{-0.3}$ & -- \\
pion$_{3}$ ${\rm log} \xi$ & $3.36^{+0.03}_{-0.03}$ & -- \\
pion$_{3}$ $\sigma ({\rm km}~{\rm s}^{-1})$ &  $294^{+7}_{-7}$ & internal turbulent velocity \\
pion$_{3}$ v (${\rm km}~{\rm s}^{-1}$) & $28^{+9}_{-6}$ & bulk velocity shift \\
pion$_{3}$ $\Omega$ & $0.23^{+0.06}_{-0.02}$ \\
\tableline
neutral Fe~K line ${\rm N}_{\rm H}$ ($10^{24}~{\rm cm}^{-2}$) & 1.6* & via ``mytorus'', fixed at Compton-thick value \\
neutral Fe~K line norm. (rel. units) & $15.1^{+0.5}_{-1.5}$ & via ``mytorus'' \\
neutral Fe~K line $r_{in}$ ($10^{3}~GM/c^{2}$) & $0.9^{+6}_{-0.2}$ & via ``spei'' \\
neutral Fe~K line $\theta$ (deg.) & $5^{+4}_{-2}$ & via ``spei'' \\
\tableline
broad Fe~XXV He-$\alpha$ norm. ($10^{42}~{\rm ph}~{\rm s}^{-1}~{\rm kev}^{-1}$) & $4.8^{+0.5}_{-0.5}$ & \\
broad Fe~XXV He-$\alpha$ E (keV) & $6.65^{+0.02}$ & range set to 6.65--6.70~keV \\
broad Fe~XXV He-$\alpha$ $\sigma$~(${\rm km}~{\rm s}^{-1}$) & $5700_{-300}$ & hit imposed upper limit \\
broad Fe~XXVI Ly-$\alpha$ norm. ($10^{42}~{\rm ph}~{\rm s}^{-1}~{\rm kev}^{-1}$) & $5.2^{+0.5}_{-0.7}$ & \\
broad Fe~XXVI Ly-$\alpha$ E (keV) & $6.97^{+0.03}_{-0.02}$ & errors incl. both imposed limits \\
broad Fe~XXVI Ly-$\alpha$ $\sigma$~(${\rm km}~{\rm s}^{-1}$) & $700_{-200}^{+100}$ & -- \\
\tableline
${\rm N}_{H,RS}$ ($10^{23}~{\rm cm}^{-2}$) & $1.1^{+0.1}_{-0.2}$ & resonant scattering, via an ``hot'' component \\
$kT_{RS}$ (keV) & $48^{+5}_{-6}$ & -- \\
$\sigma_{RS}$~(${\rm km}~{\rm s}^{-1}$) &  $2_{-2}^{+20}$ & in addition to thermal broadening \\
v$_{RS}$~(${\rm km}~{\rm s}^{-1}$) &  $-610^{-30}_{+30}$ & -- \\
$f_{cov, RS}$ & 1.0* & -- \\
\tableline
${\rm N}_{H,CE}$ ($10^{24}~{\rm cm}^{-2}$) & $3.0^{+0.3}_{-0.3}$ & warm partial covering of the the central engine\\
$kT_{CE}$ (keV) & $0.002_{-0.001}^{+0.008}$ & -- \\
$f_{cov, CE}$ & $0.92^{+0.02}_{-0.02}$ & -- \\
\tableline
${\rm N}_{H,ISM}$ ($10^{22}~{\rm cm}^{-2}$) & $8.4^{+0.1}_{-0.1}$ & column density in the ISM \\
\tableline
Flux ($10^{-10}~{\rm erg}~{\rm cm}^{-2}~{\rm s}^{-1}$, 0.0136-13.6~keV) & $3.0^{+0.3}_{-0.3}$ & the observed flux \\
Luminosity ($10^{37}~{\rm erg}~{\rm s}^{-1}$, 0.0136-13.6~keV) & $8.8^{+0.6}_{-0.6}$ & the emitted luminosity \\
\tableline
${\rm A}_{\rm Si}$ & $1.3^{+1.1}_{-0.5}$ & -- \\ 
${\rm A}_{\rm S}$ & $1.0^{+0.1}_{-0.1}$ & -- \\ 
${\rm A}_{\rm Ar}$ & $1.2^{+0.1}_{-0.1}$ & -- \\ 
${\rm A}_{\rm Ca}$ & $1.54^{+0.03}_{-0.14}$ & -- \\ 
${\rm A}_{\rm Cr}$ & $2.9^{+0.2}_{-0.4}$ & -- \\ 
${\rm A}_{\rm Mn}$ & $2.8^{+0.2}_{-0.4}$ & -- \\ 
${\rm A}_{\rm Fe}$ & $1.2^{+0.2}_{-0.1}$ & -- \\ 
${\rm A}_{\rm Ni}$ & $1.4^{+0.2}_{-0.2}$ & -- \\ 
\tableline
${\rm C}/\nu$ & 3891/2933 & --\\
\tableline
\end{tabular}
\vspace*{\baselineskip}~\\
\end{center} 
\tablecomments{Spectral model parameters values and $1\sigma$ errors.  The time-averaged spectrum was fit after binning using the ``optimal'' binning algorithm \citep{kaastra2016}, minimizing a Cash statistic.  Parameter values marked with an asterisk were held constant in the fit.  The disk and power-law components from the central engine are listed first, followed by the photoionized emission zones that they illuminate (three pion zones), neutral Fe~K emission line parameters (modeled with ``mytorus'' and ``spei''), then broadened lines modeled with Gaussians, followed by a set of internal absorption and scattering columns.  The continuum is covered with one column of warm gas via a ``hot'' absorption component. The line emission regions are covered by a hot resonant scattering zone.  The entire model is covered by a neutral column in the ISM.  Please see the text for a full description of the complex interplay between the central engine, emission regions, and absorption and scattering zones.  The abundance constraints that  the follow the other parameters were jointly determined by linking their values between the photoionized emission zones. } 
\end{scriptsize}
\end{table}

\clearpage

\begin{figure}[t]
    \centering
     \includegraphics[width=1.0\columnwidth]{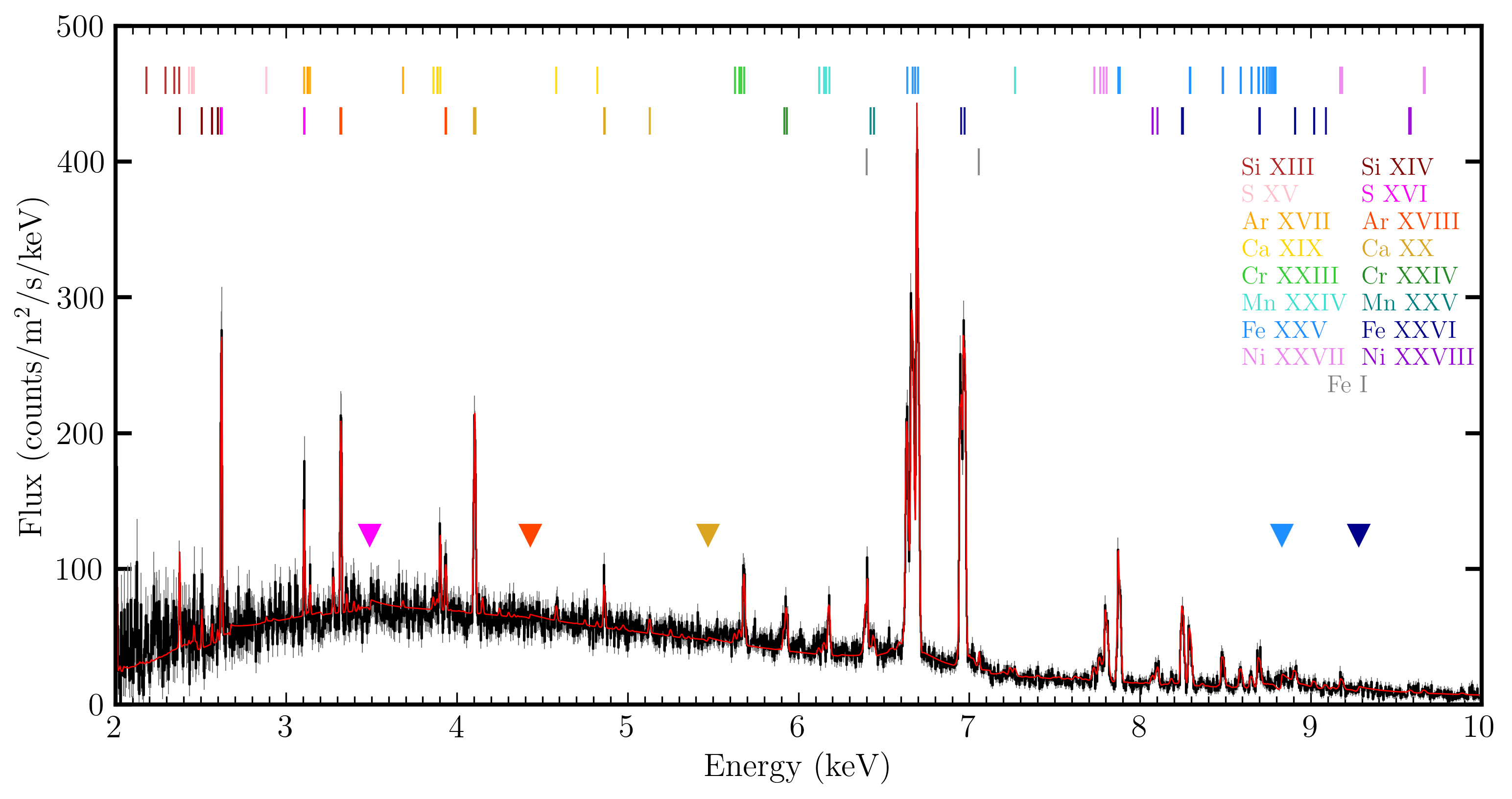}
   \caption{The XRISM/Resolve spectrum of GRS~1915$+$105 in its
     obscured state.  The model in red is the best-fit model detailed
     in Table 1.  It consists of a highly obscured disk plus power-law
     continuum and three photoionized emission zones in the outer
     accretion disk, weakly shaped by resonant scattering.  The
     strongest lines are from He-like and H-like charge states of Si,
     S, Ar, Ca, Cr, Mn, Fe, and Ni.  Solid vertical lines mark the lab
     energies of the observed He-like (top row) and H-like transitions
     (bottom row).  A weak, narrow, neutral Fe~K$_{\alpha}$ emission
     line is clearly detected at 6.4~keV, signaling the presence of
     cold gas at intermediate disk radii.  Several RRC features are
     marked with triangles, including from He-like and H-like Fe at
     8.8 and 9.3~keV, respectively.  The data are binned using the
     ``optimal'' binning algorithm.}
   \label{fig:full}
\end{figure}

\begin{figure}[t]
    \centering
     \includegraphics[width=1.0\columnwidth]{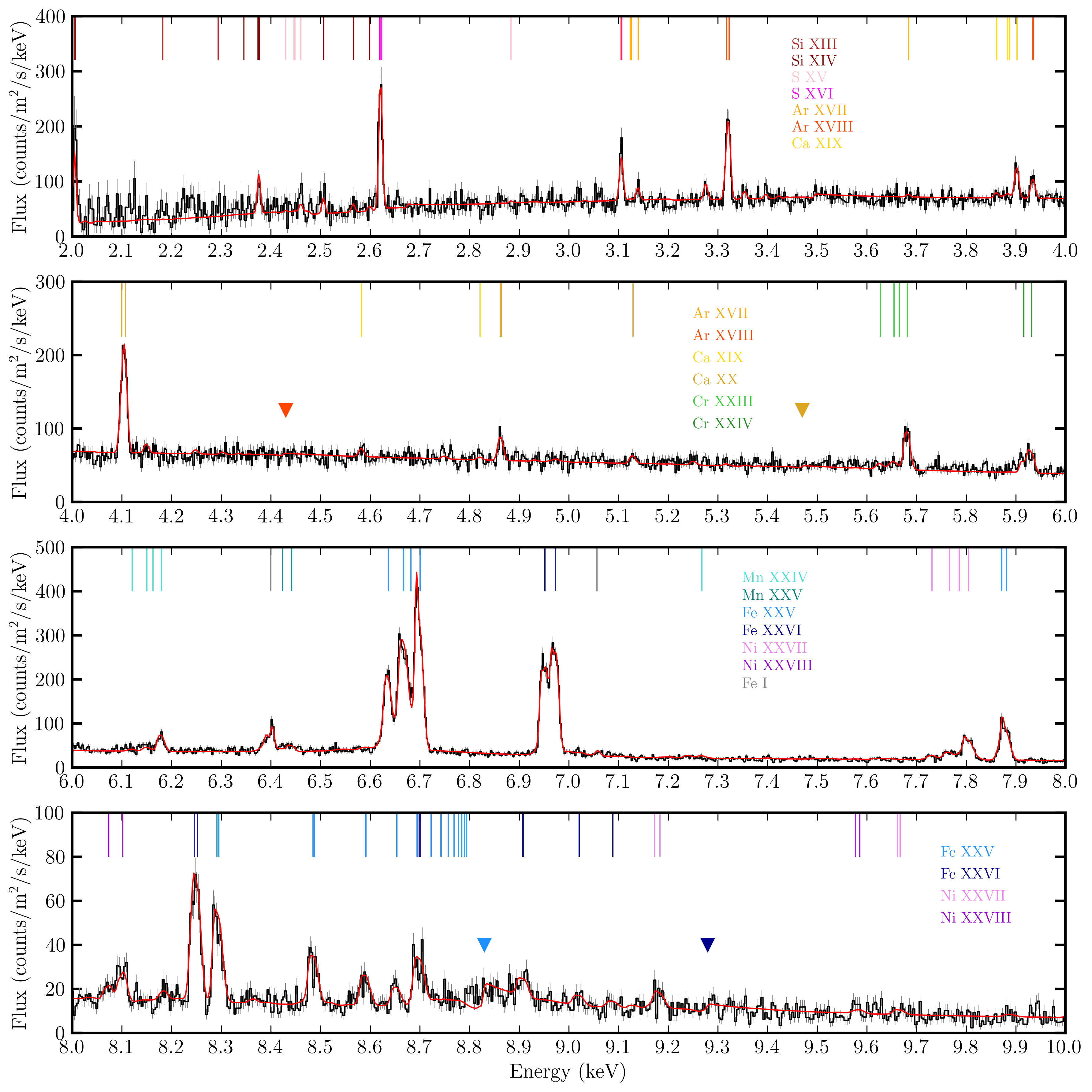}
   \caption{The XRISM/Resolve spectrum of GRS~1915$+$105, shown in
     2~keV segments.  The model in red is the best-fit model detailed
     in Table 1.  Solid vertical lines mark the lab energies of the
     observed He-like and H-like lines, as well as neutral Fe.  The
     red-shift of the bulk of the emitting gas relative to lab
     wavelengths is evident, especially in the strongest Fe~XXV and
     Fe~XXV lines.  RRC features are marked with triangles.}
   \label{fig:multi}
\end{figure}

\begin{figure}[t]
    \centering
     \includegraphics[width=1.0\columnwidth]{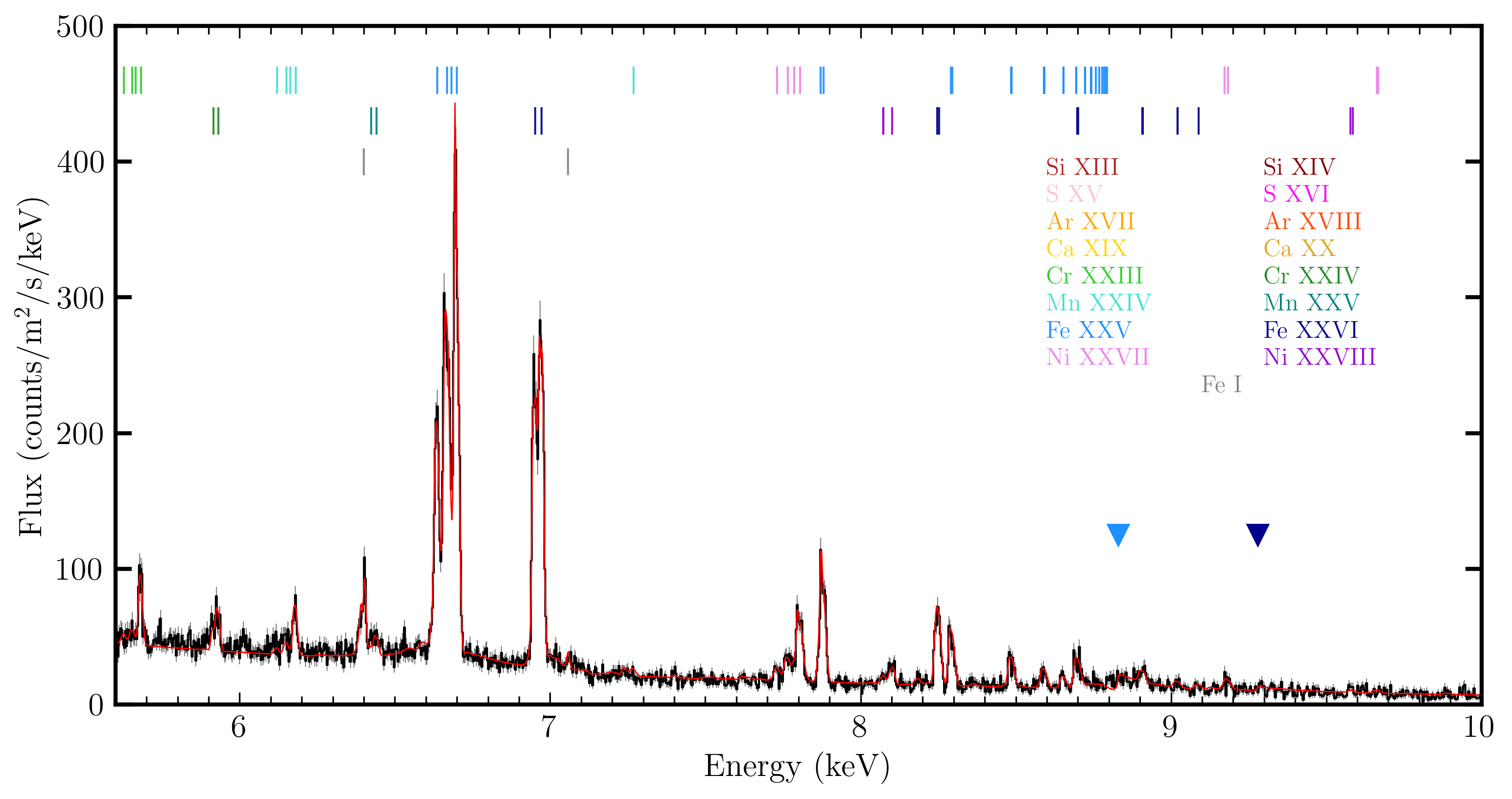}
   \caption{The XRISM/Resolve spectrum of GRS~1915$+$105, in the band
     containing He-like and H-like lines from Cr, Mn, Fe, and Ni. The
     model in red is the best-fit model detailed in Table 1.  Solid
     vertical lines mark the lab energies of the observed He-like (top
     row) and H-like transitions (bottom row).  RRC features are
     marked with triangles.}
   \label{fig:fek}
\end{figure}

\begin{figure}[t]
    \centering
     \includegraphics[width=1.0\columnwidth]{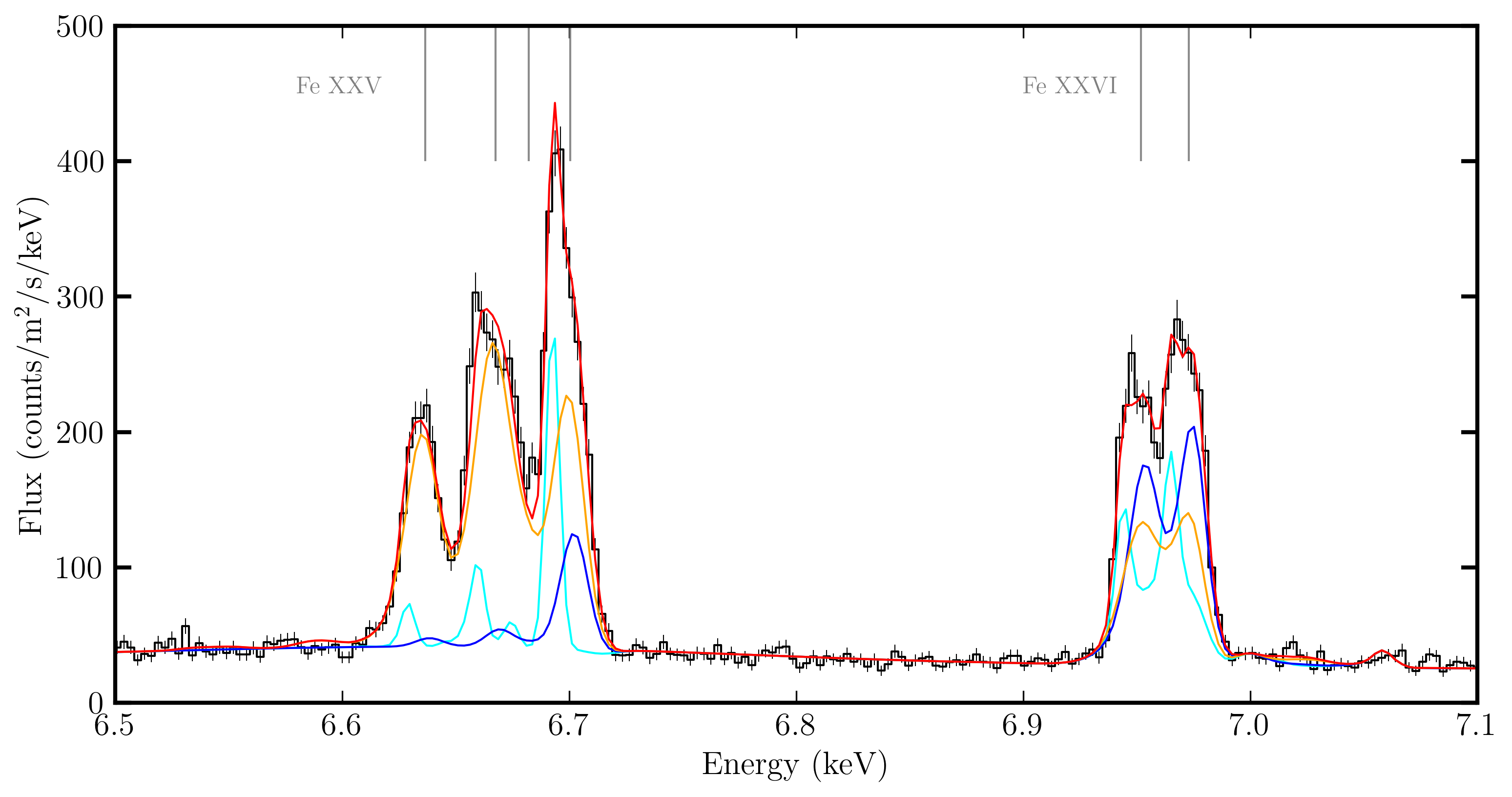}
   \caption{The XRISM/Resolve spectrum of GRS~1915$+$105 in the
     vicinity of the strongest emission line complexes (He-like Fe
     XXV, centered at 6.7~keV, and H-like Fe~XXVI centered at
     6.97~keV).  The total best-fit model is shown in red, with
     individual photoionized components shown in blue (pion$_{1}$),
     cyan (pion$_{2}$) and orange (pion$_{3}$).  In these and other
     lines from adjacent charge states, it is clear that multiple
     components are needed to describe He-like and H-like line fluxes,
     and detailed line shapes.  It is evident that the lines are
     red-shifted with respect to their laboratory energy values,
     marked in gray at the top of the frame.  Note that the component
     colors adopted in this plot are not connected to the elemental
     color scheme in other plots.}
   \label{fig:comps}
\end{figure}

\begin{figure}[t]
    \centering
     \includegraphics[width=1.0\columnwidth]{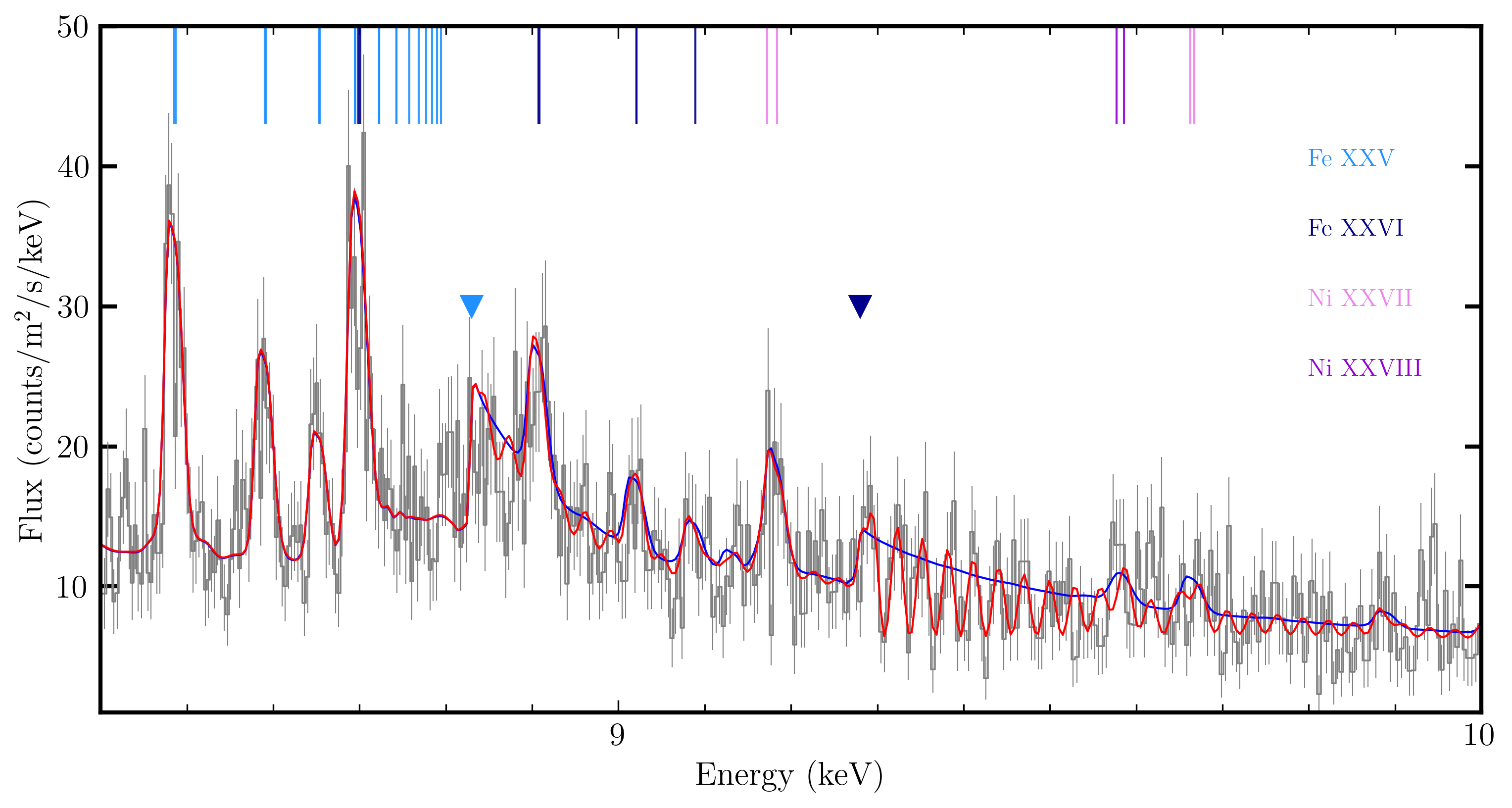}
   \caption{The XRISM/Resolve spectrum of GRS~1915$+$105 in the band
     containing Fe~XXV and Fe XXVI~RRC features (marked with
     triangles).  The best-fit model in other figures is now shown in
     blue.  The Fe XXVI~RRC is clearly not smooth, as expected if
     recombining electrons are thermally distributed; instead,
     oscillatory structure with a spacing of 30~eV is evident. This
     structure is nominally consistent with recombination from Landau
     levels in a highly magnetized corona ($B = 2.5\times
     10^{9}$~Gauss; this model is shown in red).  Alternatively, the
     lines could arise in a $v=0.34c$ flow wind, wherein iron retains
     intermediate charge states.  Both explanations are
     unsatisfactory.}
   \label{fig:mag}
\end{figure}

\begin{figure}[t]
    \centering
     \includegraphics[width=1.0\columnwidth]{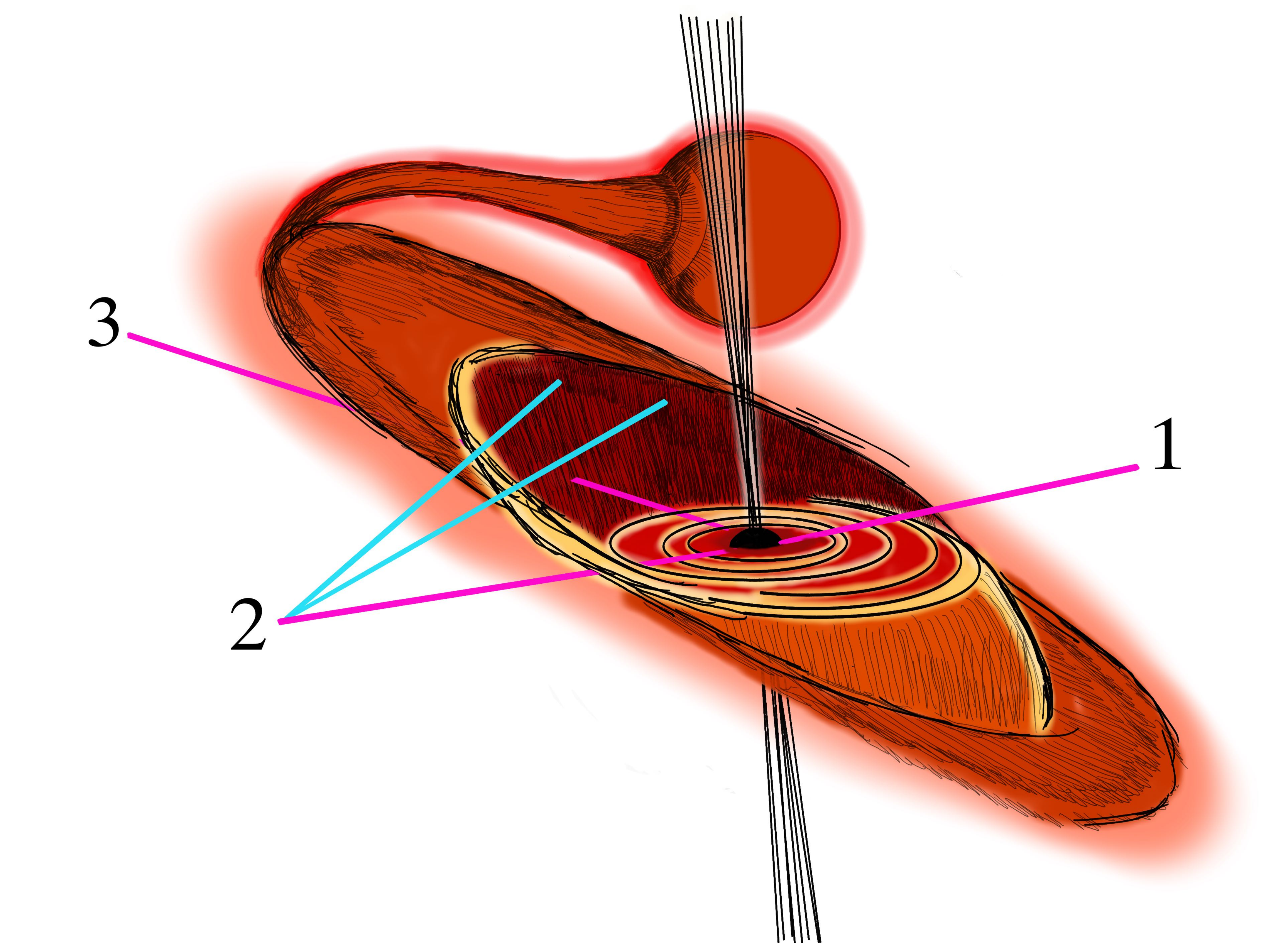}
   \caption{A diagram of a warped disk geometry that might lead to the
     spectrum observed from GRS~1915$+$105 with Resolve.  Position 1
     sees the central engine, only obscured by disk winds.  At
     Position 2, the central engine is blocked by the warp in the
     outer disk, but emission from the inner face of the warp 
     is visible.  At Position 3, the central engine and irradiated
     face of the disk are both blocked.  Note that the warp period is much slower than the orbital period of the system; it is not locked in orbital phase.}
   \label{fig:schematic}
\end{figure}

\end{document}